\documentclass[prl,fleqn,nofootinbib,showpacs,twocolumn]{revtex4}
\usepackage{amsmath}
\usepackage{mathrsfs}
\usepackage{amsfonts}
\usepackage[dvips]{graphicx}
\newcommand{\eq}{\begin{equation}}
\newcommand{\xeq}{\end{equation}}
\newcommand{\eqq}{\begin{equation*}}
\newcommand{\xeqq}{\end{equation*}}
\newcommand{\ear}{\begin{eqnarray}}
\newcommand{\xear}{\end{eqnarray}}

\begin{document}
\title{Enhanced Macroscopic Quantum Tunneling in Bi$_2$Sr$_2$CaCu$_2$O$_{8+\delta}$\\Intrinsic Josephson Junction Stacks}
\author{X.Y. Jin}
\author{J. Lisenfeld}
\author{Y. Koval}
\author{A. Lukashenko}
\author{A.V. Ustinov}
\author{P. M\"uller}
\affiliation{Physikalisches Institut III, Universit\"at
Erlangen-N\"urnberg, Erwin-Rommel-Strasse 1, D-91058 Erlangen,
Germany}
\date{\today}

\begin{abstract}
We have investigated macroscopic quantum tunneling in
Bi$_2$Sr$_2$CaCu$_2$O$_{8+\delta}$ intrinsic Josephson junctions
at millikelvin temperatures using microwave irradiation.
Measurements show that the escape rate for uniformly switching
stacks of N junctions is about $N^2$ times higher than that of a
single junction having the same plasma frequency. We argue that
this gigantic enhancement of macroscopic quantum tunneling rate in
stacks is boosted by current fluctuations which occur in the
series array of junctions loaded by the impedance of the
environment.
\end{abstract}


\maketitle\rm

After 20 years since its discovery, macroscopic quantum tunneling
(MQT) in Josephson junctions remains a fascinating phenomenon
which attracts interest of a broad physics community. MQT was
first observed in Nb Josephson junctions at very low temperatures
 \cite{Voss} and has been used to study energy level
quantization by microwave absorption \cite{Martinis1}. More
recently, so-called phase qubits based on MQT in current-biased
Josephson junctions have been reported as very promising hardware
for quantum information processing \cite{Martinis4, MD1,
Martinis5}.

Due to their \textit{d}-wave order parameter symmetry
\cite{Tsuei}, cuprate high-T$_c$ superconductors were initially
regarded unsuitable for MQT experiments. However, it has been
argued that although there are nodes in the d-wave order
parameter, MQT should not be suppressed completely in high-T$_c$
Josephson junctions \cite{Japs1}. Meanwhile, recent measurements
on YBa$_2$Cu$_3$O$_{7-\delta}$ grain boundary junctions show the
solid evidence of MQT \cite{Delsing1}.

On the other hand, intrinsic Josephson junctions (IJJs) in layered
high-T$_c$ superconductors \cite{Paul1, Paul2} are rather
attractive candidates for MQT experiments. IJJs in
Bi$_2$Sr$_2$CaCu$_2$O$_{8+\delta}$ are formed by pairs of CuO$_2$
double planes, separated by a Bi$_2$O$_3$ insulating layer. They
have a much higher Josephson coupling energy than grain boundary
junctions and a better homogeneity, as they are located inside a
more or less perfect single crystal. Moreover, intrinsic junctions
exhibit current transport along the $c$-axis direction, i.e.
perpendicular to the copper oxide layers, where the nodes of the
$d_{x^2-y^2}$ order parameter should not affect MQT at all. With
the recent invention of a double-sided fabrication technique
\cite{Wang1}, one can avoid heating the junctions by contact
resistance. This makes MQT in IJJs practically attainable.
Recently, MQT has been observed on a single junction in
Bi$_2$Sr$_2$CaCu$_2$O$_{8+\delta}$ IJJs \cite{Japs3, Wang2}. The
temperature $T^*$ of the crossover between thermal and quantum
escape was reported to be rather high.

In this Letter, we present an experimental study of MQT in IJJs
using microwave spectroscopy. We demonstrate that the unique
uniform array structure of intrinsic Josephson junction stacks
causes an enormous enhancement of the tunneling rate. We argue
that this enhancement can be caused by current fluctuations in the
stack.

The samples were fabricated using the standard double-sided ion
beam etching technique \cite{Wang1}. The stack height ranges from
approximately 7.5 nm to 15 nm, i.e. the number of junctions N in
our IJJ series arrays is between 50 and 100. The junction area
varied for different stacks from 1$\times$2 $\mu$m$^2$ to
2$\times$3 $\mu$m$^2$. Critical current densities $j_c$ were
between 0.7 - 2.8 kA/cm$^2$. Sample parameters are summarized in
Table 1. Current-voltage (I-V) characteristics were measured with
a standard four-terminal configuration using current biasing.
Experiments were performed in a $^3$He cryostat with a minimum
temperature of 300 mK and in a $^3$He/$^4$He dilution refrigerator
with a base temperature of 10 mK.
\begin{figure}[b]
\includegraphics[width=8.5cm]{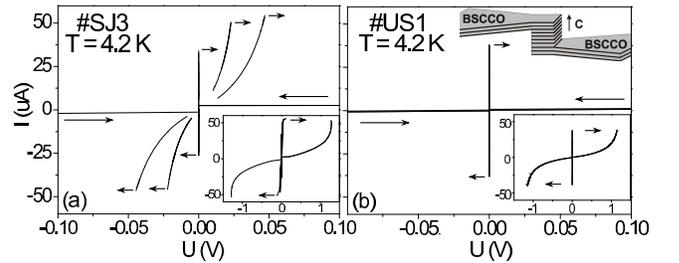}
\caption{I-V characteristics: (a) One-by-one switching of three
junctions in a stack, $\sharp$SJ3; (b) uniform switching of the
whole stack, $\sharp$US1. The insets show the full-range I-V
curves for both samples. The sample geometry is sketched in the
upper right corner.}
\end{figure}

We investigated two different types of samples. Their typical I-V
characteristics are presented in Fig. 1. In the first case, at
least one junction had a critical current significantly lower than
those of the rest of the stack. This was achieved either by
exposing the unprotected side of the crystal to ambient
atmosphere, resulting in a reduced critical current density, or by
over-etching resulting in a reduced cross-sectional area. When
increasing the current, we observed voltage jumps from one branch
to the next in steps of approximately 25 mV (Fig.\ 1a). Due to the
large difference of the critical current, it is always the same
junction that switches to its resistive state first. We call this
behavior "single-junction switching" and denote this type of
samples as the $\sharp$SJ series. In the second case, the stack is
so homogeneous that upon \emph{increasing} the current, we never
observed any stable states between zero voltage and 1.14 V, where
all junctions are in the resistive state (Fig.\ 1b). We call that
behavior "uniform-stack switching" and denote this type of samples
as the $\sharp$US series. Nevertheless, we find that many branches
corresponding to different numbers of resistive junctions are
still there. They can be easily traced up from the return curve of
the resistive state\cite{Paul1, Paul2, Wang1}.
\begin{figure}[t]
\includegraphics[width=7.5cm]{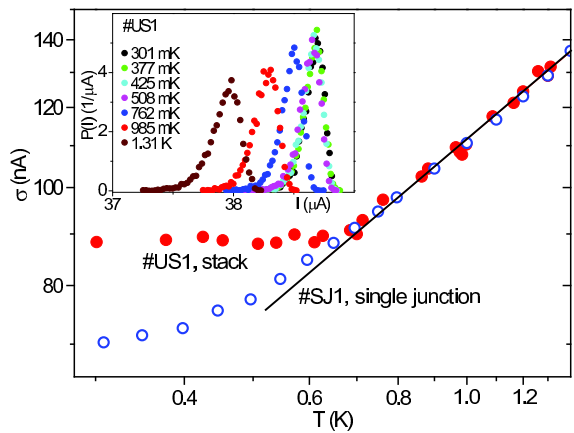}
\caption{Temperature dependence of the standard deviation of the
switching current from the zero-voltage state. Solid circles:
sample $\sharp$US1; open circles: sample $\sharp$SJ1. The straight
line is a linear fit to all data points above 700 mK. The inset
shows switching current distributions for $\sharp$US1. For better
comparison, the $\sharp$US1 data were divided by a factor of
1.09.}
\end{figure}
\begin{table}[b]
\caption{Sample parameters measured.}
\begin{tabular}{lclcclc}
$\sharp$&size($\mu$m$^2$)&$I_s$($\mu$A)&$T^*$(mK)&$f_p^0$(GHz)&$\gamma_s$(\%)&N\\
\hline{}SJ1&2$\times$3&57.7&300&120&99.0&1\\
SJ2&2$\times$3&162.5&450&180&99.2&1\\
SJ3&2$\times$3&67.4&320&135&98.9&1\\
US1&1$\times$3&38.2&700&138&96.5&46\\
US2&2$\times$2&31.0&500&126&96.3&42\\
US3&2$\times$3&57.2&550&140&97.2&50\\
US4&2$\times$3&65.5&620&150&97.3&$\sim$100
\end{tabular}\end{table}

Distributions of the switching current, $I_s$, at which the
junctions escape from the zero voltage state \cite{Fulton1} were
measured using a high-resolution ramp-time based setup
\cite{Ustinov2}. The bias current was ramped up with a rate of 200
mA/s. After detecting a switching event by a voltage threshold of
20 $\mu$V, the current was switched to zero within less than 10
$\mu$s in order to avoid heating effects. Monitoring the voltage
on a fast oscilloscope for $\sharp$US-type samples showed that the
current decrease was fast enough so that any stack always switched
to its first resistive branch, i.e. to a state with one resistive
junction. Based on this fact we emphasize that regarding
dissipation both types of samples were measured under
\emph{exactly} the same experimental conditions. Switching current
distributions were measured at temperatures between 20 mK and 12 K
using a repetition rate of 600 Hz. The switching current
statistics was determined from 20,000 to 60,000 switching events.
The switching probability P(I) shown in the inset of  Fig.\ 2 is
defined as the number of switching events per $\mu$A normalized to
the total number of events. The standard deviation $\sigma$ of the
switching current was determined from the width of the P(I) curve.
\begin{figure}[b]
\includegraphics[width=8cm]{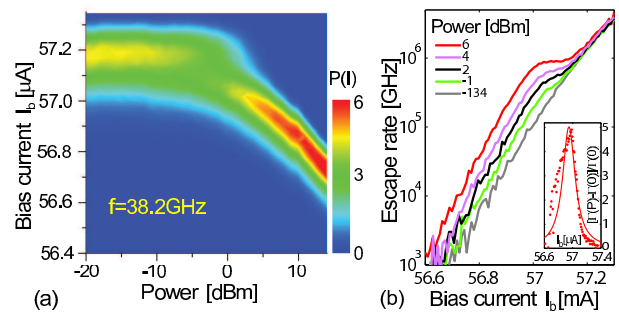}
\caption{(a) Density plot of switching current vs. microwave power
for two-photon absorption in sample $\sharp$US3 at 30 mK. The
switching probability P(I) is color coded according to the bar on
the right-hand side. (b) The corresponding enhancement of the
escape rate at different microwave powers. The inset shows the fit
of the 6 dBm enhancement curve with a Lorentzian.}
\end{figure}
Fig. 2 shows $\sigma$ as a function of temperature for samples
$\sharp$SJ1 and $\sharp$US1. The saturation of $\sigma$(T) at low
temperatures corresponds to a crossover from thermal activation to
MQT. For sample $\sharp$SJ1 the saturation is not complete in
Fig.~2. This is due to the fact that this experiment was done in a
$^3$He cryostat where the lowest temperature was 300 mK. We
verified complete saturation in our experiments in the dilution
refrigerator. Nevertheless, one may still see in Fig. 2 that the
crossover temperature $T^*$ of the single-junction sample
$\sharp$SJ1 is about 300 mK, while the uniform-switching sample
$\sharp$US1 shows $T^*$ of about 700 mK. This discrepancy can not
be fully accounted by the difference in the critical currents.

At temperatures below $T^*$, we measured the switching current
distributions under microwave radiation in the frequency range
between 10 and 40 GHz. Such measurements allow to determine the
plasma frequency $\omega_p^0$ and the absolute value of the
fluctuation-free critical current $I_c$ directly \cite{Martinis1}.
Here, microwave spectroscopy serves as a tool to determine the
energy level separation in the quantum regime. It should be
noticed that quantum transitions between levels cannot be easily
distinguished from classical plasma resonance peaks using just
spectroscopy data \cite{theory0}. However, below the crossover
temperature $T^*$ quantum fluctuations dominate thermal ones and
quantum mechanics is appropriate for describing the system. Since
the zero-bias plasma frequency $f_p^0=\omega_p^0/2\pi$ for the
samples was well above 100\ GHz, we used multi-photon
\cite{Ustinov1} rather than single-photon absorption in our
measurements.

As the microwave power is increased, the P(I) distribution becomes
double-peaked. The microwave-induced peak at lower currents
corresponds to a plasma resonance in the junction. In the quantum
picture, this peak is interpreted as tunneling from
highly-populated energy level(s). Figure~3(a) shows a density plot
of the switching current distribution of a uniform-switching stack
versus microwave power at 38.2 GHz. Figure~3(b) shows the
corresponding enhancement of the escape rate. The double-peaked
P(I) distribution develops at a microwave power of about -1 dBm
referred to the top of the cryostat. The two-photon resonance
current peak appears at about 57.0 $\mu$A. It should be noted that
the width of the resonance peak is smaller than the distribution
width at zero microwave power, which indicates that the standard
deviation of P(I) measured without microwaves is not limited by
current noise in our setup.
\begin{figure}[t]
\includegraphics[width=7cm]{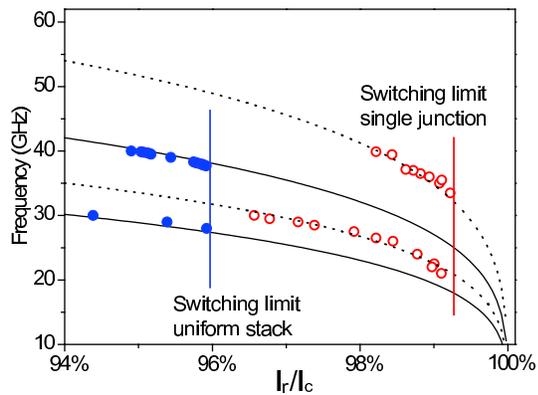}
\caption{Applied microwave frequency versus normalized resonance
current for sample $\sharp$SJ2 (open circles) and $\sharp$US3
(solid circles) measured at 20 mK. The data are fitted to Eq.(1)
using n=2 (upper dotted line and upper solid line) and n=3 (lower
lines). The vertical lines mark the maximum switching current of
the samples.}
\end{figure}
The resonance current $I_r$ is defined as the position of the
resonant peak when both peaks have equal amplitudes. The results
at different microwave frequencies are summarized in Fig.~4. The
data are fitted to \cite{Martinis1}
\eq\omega_p=\frac{1}{n}\omega_p^0(1-\gamma^2)^{1/4}\:,\xeq where
$\omega_p$ is the plasma frequency of the junction at the
normalized bias current $\gamma$, $\omega_p^0$ is the plasma
frequency at zero bias, and $n$ is the number of photons taking
part in the absorption process. $\gamma=I_b/I_c$ is given in
normalized units, where $I_b$ is the bias current and $I_c$ is the
fitted fluctuation-free critical current. The data in Fig.~4 show
the best fit to Eq.~(1) by assuming two- and three-photon
absorption. At lower frequencies and high powers we observed
multi-photon peaks up to $n=6$ (not shown). From the fits we
obtain $f_p^0=\omega_p^0/2\pi=150$\ GHz (sample $\sharp$US3), and
$f_p^0=180$\ GHz (sample $\sharp$SJ2). The fitted $f_p^0$ of the
other samples are shown in Table 1. The obtained plasma
frequencies provide an estimate for the junction capacitance per
unit area C=$j_c/(2\pi\Phi_0f_p^0)$, where $\Phi_0$ is the
magnetic flux quantum. Our measurements yield C=3.9 $\mu$F/cm$^2$
and $\varepsilon_r$=5.3, which conforms well with the value of
$\varepsilon_r$=5 obtained in earlier work \cite{Paul4}.

From the results shown in Fig.\ 4 and Table\ 1, one may notice
that the normalized switching current $\gamma_s=I_s/I_c$ of the
uniformly switching stacks is significantly lower than that of the
single junction switching samples. The typical value of $\gamma_s$
for single-junction samples is about $99\%$ of the
fluctuation-free critical current $I_c$, whereas for
uniformly-switching stacks it is only $96\%$. The switching range
of the single-junction samples is fairly similar to that of Nb
junctions with comparable critical current densities, where
$\gamma_s$ is also close to $99\%$ \cite{Ustinov1}. On the
contrary, the normalized switching bias current $\gamma_s$ of
uniformly switching stack samples is about 3\% smaller. This
discrepancy can explain the difference in $T^*$ for $\sharp$SJ and
$\sharp$US samples. The plasma frequency $\omega_p^s$ at the
switching current is given by Eq.~(1) with $\gamma=\gamma_s$. For
a single junction, the temperature of the crossover from thermally
activated escape to MQT is expected \cite{Grabert} to be
$T^*\approx\hbar\omega_p^s/2\pi k_B$. For uniformly-switching
samples, the premature switching at lower $\gamma_s$ leads to the
high $\omega_p^s$ and thus to an enhanced $T^*$.

In the quantum regime, a single Josephson junction escapes through
the energy barrier with an escape rate $\Gamma$ determined by
\cite{Leggett}
\eq\Gamma=\frac{\omega_p}{2\pi}\left(\frac{864U_0\pi}{\hbar\omega_p}\right)^{1/2}\exp\left(-\frac{36}{5}\frac{U_0}{\hbar\omega_p}\right)\xeq
where $\omega_p$ is given by Eq.~(1),
$U_0=E_J\frac{4\sqrt{2}}{3}(1-\gamma)^{3/2}$ is the height of the
potential barrier at $\gamma\lesssim 1$, and $E_J=\Phi_0 I_c/2\pi$
is the Josephson coupling energy. In Fig.~5 we plot Eq.~(2) using
the values of $\omega_p^0$ and $I_c$ in Table 1 for samples
$\sharp$US1, $\sharp$US4 and $\sharp$SJ3, respectively (solid
lines). The interconnected dots were determined from the measured
P(I) histograms. For single junction sample $\sharp$SJ3, one can
see that experiment agrees well with the theory. However, there is
a huge difference between theory and experiment for both uniformly
switching samples.

\begin{figure}[t]
\includegraphics[width=7.5cm]{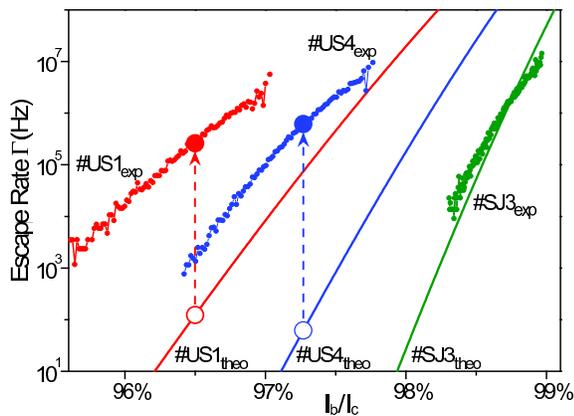}
\caption{Escape rate versus bias current for uniformly switching
stack samples $\sharp$US1 and $\sharp$US4 and single-junction
switching sample $\sharp$SJ3. The solid lines are calculated by
the single-junction quantum escape theory (2). The points
interconnected by lines are extracted from the measured P(I)
distributions. The vertical-arrows indicate the enhancement of the
escape rate (see text).}
\end{figure}

We used the mean switching bias $\gamma_s$ of each sample to mark
the intersection of $\gamma_s$ with the theoretical curves in
Fig.~5 by large open dots. According to the theory, these dots
should correspond to the most probable escape current. One can see
that the intersections of $\gamma_s$ with the experimental curves
show actual escape rates higher by factor of roughly 2000
($\sharp$US1) and 10000 ($\sharp$US4). This difference is marked
by the vertical arrows in Fig.~5. Samples listed in Table 1 have
different sizes and $j_c$'s. We have measured them in different
setups and at different temperatures. Experiments for \emph{all}
single junction samples well agree with theory (2). On the other
hand, for \emph{all} uniformly switching samples we always found a
huge difference between the measured data and single-junction
model given by Eq.~(2). We therefore conclude that the single
junction tunneling theory is not applicable to the uniform
switching stacks.

The value of $N$ in $\sharp$US samples can be determined by
counting the number of resistive branches in the $I$-$V$
characteristics. Assuming that the single junction escape rate is
enhanced by a factor of $N^2$, one can renormalize the peak escape
rate of all $\sharp$US samples. In Fig.~5 we have chosen to show
samples $\sharp$US1 with $N=46$ and $\sharp$US4 with $N=100$. We
find that the $N^2$ corrected values shown by large solid dots
conform pretty well to the experimental data. Moreover, our data
for samples $\sharp$US2 and $\sharp$US3 (see Table 1) also match
this $N^2$ correction.

If there would be no interaction between $N$ identical junctions
in a current-biased series array, the escape rate $\Gamma$ (the
probability that at least one junction switches at a given bias
current) should be merely enhanced by a factor $N$ with respect to
a single junction. If we assume that switching of any junction in
the array is triggered, in addition to its own fluctuations, by
its nearest neighbors in the array (e.g., via charge coupling
through the shared CuO$_2$ double-planes), then we should get an
enhancement by factor of about $3N$. However, experiments imply
that the enhancement of the escape rate in stacks is proportional
to $N^2$. This is possible when there is an interaction between
\emph{any pair} out of the $N$ junctions. Such interaction occurs
when the stack of junctions connected in series is loaded in
parallel by a relatively low impedance. Fluctuations of the phase
difference of a single junction change its parametric Josephson
inductance and thus the total inductance of the array. The
external bias current is split-up between the array and the
external impedance, which can be regarded as the impedance of the
environment at the plasma frequency. The bias current flowing
through the array thus changes under fluctuations in any junction.
A specific analysis of this model goes beyond the scope of this
experimental paper.

In conclusion, we have found a drastically enhanced escape rate of
macroscopic quantum tunneling in uniformly switching
Bi$_2$Sr$_2$CaCu$_2$O$_{8+\delta}$ intrinsic Josephson junction
stacks. This enhancement adds a factor of approximately $N^2$ to
the quantum escape rate of a single Josephson junction. This can
be caused by large quantum fluctuations due to interactions among
the N junctions and results in a significant increase of the
crossover temperature $T^*$ between the thermal activation regime
and quantum tunneling.

We thank Dr. H. B. Wang for providing high quality
Bi$_2$Sr$_2$CaCu$_2$O$_{8+\delta}$ crystals and for helpful
discussions. This work was supported by Deutsche
Forschungsgemeinschaft.

\end{document}